\documentstyle[epsf,aps,preprint,floats,tighten]{revtex}

\begin{document}
\preprint{\vbox{\hbox{CERN-TH/97-361}
                \hbox{hep-ph/9712343}
                \hbox{November 1997}
                }
}

\title{The Role of Scalar and Pseudoscalar Fields in Determining 
Nucleosynthesis Bounds on the Scale of Supersymmetry Breaking}

\author{Tony Gherghetta\footnote{Tony.Gherghetta@cern.ch}}
\address{Theory Division, CERN, CH-1211 Geneve 23, Switzerland}

\maketitle

\begin{abstract}
The effect of spin-0 goldstino superpartners is considered
on the nucleosynthesis bounds arising when a superlight gravitino appears 
as an effective massless neutrino species. When the scalar and pseudoscalar
superpartners are relativistic they will decouple at much later times 
than the goldstino and consequently will be the dominant effect when 
obtaining a nucleosynthesis bound on the scale of supersymmetry breaking. 
Assuming that the scalar and pseudoscalar fields decouple at a 
temperature no later than ${\cal O}(100)$ MeV, then typically the scale 
of supersymmetry breaking $\sqrt{F}\gtrsim 60$ TeV. This corresponds 
to a lower bound on the gravitino mass $m_{3/2}\gtrsim 1$ eV.
\end{abstract}

\newpage

\section{Introduction}

If supersymmetry is realised in nature then it is presumably 
broken at some energy scale and subsequently transmitted indirectly
to the low energy observable mass spectrum \cite{spm}. While naturalness 
arguments essentially constrain the amount of supersymmetry breaking in 
the low energy mass spectrum, the scale at which supersymmetry is broken 
remains largely undetermined. In general, 
realistic supersymmetry breaking scenarios can be parametrised by a spurion
chiral supermultiplet $(\phi,\psi,F)$ in which a nonzero vacuum expectation
value for the F-term gives rise to the spontaneous breaking of supersymmetry 
at the scale $\sqrt{F}$. The fermion, $\psi$ represents the massless 
Nambu-Goldstone fermion which becomes part of the massive gravitino after 
the superHiggs effect. This fermion is referred to as the goldstino 
$\tilde{G}$. The gravitino mass $m_{3/2}$ is determined purely by 
gravitational interactions and is fixed by the relation
$F=\sqrt{3}m_{3/2}M_P$, where the reduced Planck mass $M_P=\sqrt{8\pi 
G_N}=2.4\times 10^{18}$GeV is related to the gravitational Newton constant 
$G_N$. When the gravitino mass $m_{3/2}\rightarrow 0$, the goldstino
becomes the dominant component of the gravitino during interactions.

The mass of the spin-0 superpartner $\phi$, is however not fixed
and can in general depend on nonminimal 
terms in the Kahler potential as well as terms in the superpotential. 
When the complex scalar is decomposed into real fields 
$\phi=(S+iP)/\sqrt{2}$, we need to consider the scalar mass $m_S$ and 
pseudoscalar mass $m_P$ as additional parameters to the supersymmetry 
breaking scale $F$. In this framework we will
be interested in placing a lower bound on the scale of supersymmetry
breaking from the effect of a superlight gravitino together with its
scalar superpartners contributing as extra effective neutrino species
during the nucleosynthesis epoch.

The effect of a superlight gravitino in the early universe was considered
in a previous calculation \cite{tg}. There it was assumed that the
spin-0 superpartners were essentially massless ($m_S,m_P \ll T$, where $T$
is the temperature during the nucleosynthesis epoch) and that they
decouple before the superlight gravitino. More recently, there have 
appeared additional papers \cite{bfzI,bfzII,lp} in which the general 
question of determining
an effective gravitino Lagrangian have been addressed. In addition
these references investigated the case of massive superpartner
masses $m_S,m_P \gg T$ and noticed qualitatively different behaviour for
scattering cross sections than the superlight scenario. However they
did not address the question of superlight scalar superpartners during
nucleosynthesis.

In this work we will consider the effect of the scalar superpartners 
in determining the nucleosynthesis bounds on the supersymmetry breaking 
scale $F$ and in particular consider the case of superlight masses for 
the scalars. It will be shown that $S$ and $P$ decouple much later than 
the goldstinos, so that the scalar and pseudoscalar contributions 
actually strengthen the bound
considered in \cite{tg}, when $m_S$ and $m_P$ are superlight as well.
While we provide a more complete study of the effect of the scalar and
pseudoscalar fields during nucleosynthesis, the result for superlight 
$S, P$ masses will agree with \cite{gmr}.
In section 2 we will consider the scattering cross sections for
$S,P$ and $\tilde{G}$ which are relevant during the nucleosynthesis
epoch. This will then allow us, in section 3, to obtain an expression 
for the corresponding decoupling temperature. We will see that this leads 
to a bound on the supersymmetry breaking scale, $F$ and consequently
a lower bound on the gravitino mass $m_{3/2}$. The conclusion and final 
comments appear in section 4.

\section{Scattering cross sections}

We will assume that all gravitino interactions are dominated by a 
superlight goldstino in determining the gravitino scattering
cross sections. The associated scattering cross sections of the
scalar superpartners will also be considered when they are superlight as well.
We will make the usual assumption that a massless and weakly interacting
particle decouples when the relic abundance is frozen.
Thus it will be sufficient to consider processes that change particle number.
The scattering cross sections can be obtained once all the effective 
interactions between the relevant particle species is known. While the full
supergravity formalism was employed in \cite{brI,tg}, it will be simpler
instead to use the effective Lagrangian approach adopted by \cite{bfzI} 
where the effects of gravity are formally decoupled in the limit 
$M_P\rightarrow\infty$ while keeping $F$ fixed
\footnote{An alternative approach using a nonlinear realisation of the 
supersymmetry algebra can also be used \cite{lp}.}.
During the nucleosynthesis era the particle species that are assumed to be in 
thermal equilibrium besides $S,P$ and $\tilde{G}$ will be the photon 
$(\gamma)$, three families of massless neutrinos $(\nu_i)$ and fermions, 
including at least electrons $(e^\pm)$ and possibly muons $(\mu^\pm)$. 
The temperature of the universe at these times will be in the range 
${\cal O}(1-100)$ MeV.

\subsection{Goldstino annihilation}

The cross section for goldstino annihilation in the massless $S,P$ limit
was previously calculated for various channels in \cite{tg,brI}. For 
goldstino annihilation into photons the result is
\begin{equation}
\label{csGGgg}
    \sigma({\tilde G}{\tilde G}\rightarrow \gamma\gamma) =
    {1\over 64\pi} {{\tilde M}^2\over F^4} s^2,
\end{equation}
where $s$ denotes the Mandelstam variable for the total energy in the
center of mass frame and $\tilde M$ generically denotes the gaugino mass
(the effects of neutralino mixing will be ignored in this paper). The 
result (\ref{csGGgg}) is valid in the limit $m_{3/2},m_S,m_P \ll \sqrt{s}
\ll {\tilde M}$. It is also straightforward to obtain the annihilation
cross section in the limit $\sqrt{s}\ll m_S,m_P,{\tilde M}$ for which
one finds
\begin{equation}
\label{csGGggC}
  \sigma({\tilde G}{\tilde G}\rightarrow \gamma\gamma) = {1\over 640\pi}
  {s^3\over F^4}. 
\end{equation}
This expression for the annihilation cross section agrees with the
result derived by \cite{bfzI,lp}, where it was noted that the
cancellations of the leading order terms change the massless $S,P$
limit cross section (\ref{csGGgg}). 
These cancellations are obviously absent when one derives the cross section
in the massless $S,P$ limit and there
is no inconsistency with the previous literature \cite{brI,brII,tg}, in 
which the massless limit was always the underlying assumption.

Similarly, the Goldstino annihilation into fermions was calculated in
\cite{tg,bfzII}. As shown in \cite{bfzII} the cross section for annihilation
into fermions is ambiguous up to an unknown parameter which represents
the neglect of higher derivative terms in the effective Lagrangian 
approach. While this ambiguity will be seen to have no direct consequences 
on the results obtained in this paper, we will nevertheless use a minimum
value of the cross section, namely \cite{bfzIII}
\begin{equation}
\label{csGGff}
    \sigma({\tilde G}{\tilde G}\rightarrow f{\bar f}) =
    {1\over 1280\pi} {s^3\over F^4},
\end{equation}
in order to derive a model independent bound.

The above annihilation channels into photons and fermions 
are not the only possible channels. Since the goldstino superpartners
$S$ and $P$ are also assumed to be in thermal equilibrium we need to
also consider the annihilation channels ${\tilde G}{\tilde G}
\rightarrow SS,PP,SP$. The effective Langrangian governing these 
interactions is given by
\begin{equation}
\label{SPGGlag}
      {\cal L}=-{1\over 2\sqrt{2}}{m_S^2 \over F}S {\tilde G}{\tilde G}
      -{i\over 2\sqrt{2}}{m_P^2 \over F}P {\tilde G}{\tilde G} + h.c.
\end{equation}
This leads to the following Goldstino annihilation cross sections into 
the scalar superpartners $S,P$ in the limit $m_S,m_P \ll \sqrt{s}$
\begin{eqnarray}
\label{csGGSP1}
    \sigma({\tilde G}{\tilde G}\rightarrow SS) &=& {1\over 128\pi}
    {m_S^8\over F^4} {1\over s} \left[ \log\left({s\over m_{3/2}^2}\right)
    -2 \right], \\
\label{csGGSP2}
    \sigma({\tilde G}{\tilde G}\rightarrow PP) &=& {1\over 128\pi}
    {m_P^8\over F^4} {1\over s} \left[ \log\left({s\over m_{3/2}^2}\right)
    -2 \right], \\
\label{csGGSP3}
    \sigma({\tilde G}{\tilde G}\rightarrow SP) &=& {1\over 16\pi}
    {m_S^4 m_P^4\over F^4} {1\over s} \left[ \log\left({s\over m_{3/2}^2}
     \right) -2 \right].
\end{eqnarray}

Clearly in the massless limit these cross sections are significantly 
smaller than (\ref{csGGgg}) and will contribute negligibly to the overall 
goldstino annihilation rate. 

\subsection{$S$ and $P$ annihilation}

The annihilation channels available for $S$ and $P$ are similar to 
those for the goldstino, except that unlike goldstinos there are no fermion
couplings in the leading order $1/F$. Thus the effective Lagrangian 
describing their interactions will be the goldstino
Lagrangian (\ref{SPGGlag}) together with the interaction with photons 
\cite{bfzI}
\begin{eqnarray}
\label{SPgglag1}
      {\cal L}&=&{-1\over 2\sqrt{2}}{\rm Re}\left({{\tilde M}\over F}\right) 
      S F_{\mu\nu} F^{\mu\nu} + {1\over 4\sqrt{2}}{\rm Im}\left({{\tilde M}
      \over F}\right) S \epsilon^{\mu\nu\rho\sigma} F_{\mu\nu} 
      F_{\rho\sigma} \\
\label{SPgglag2}
      &&+{1\over 2\sqrt{2}}{\rm Im}\left({{\tilde M}\over F}\right) 
      P F_{\mu\nu} F^{\mu\nu} + {1\over 4\sqrt{2}}{\rm Re}\left({{\tilde M}
      \over F}\right) P \epsilon^{\mu\nu\rho\sigma} F_{\mu\nu} 
      F_{\rho\sigma}
\end{eqnarray}
In the limit that $m_S,m_P\ll \sqrt{s}$ this leads to an annihilation
cross section
\begin{equation}
\label{csSSgg}
    \sigma(SS\rightarrow \gamma\gamma) = \sigma(PP\rightarrow \gamma\gamma)
    = {5\over 64\pi} {|{\tilde M}|^4\over F^4} s,
\end{equation}
obtained via t and u-channel photon exchange.
This cross section agrees with the result obtained 
by employing the full supergravity formalism \cite{brII}.
Notice that the dependence on the Mandelstam variable $s$ is 
qualitatively different from that of (\ref{csGGgg}) and we can already 
see that $S$ and $P$ interact more strongly with the thermal plasma
when $T\lesssim{\tilde M}$. This will ultimately correspond to $S$ and 
$P$ decoupling later than the goldstino.

There are also possible annihilation channels into goldstino pairs 
due to the interactions (\ref{SPGGlag}). These are the inverse of the 
processes considered in (\ref{csGGSP1}-\ref{csGGSP3}) 
and give rise to similar cross 
sections for annihilation into goldstino pairs, namely $\sigma(SS\rightarrow 
{\tilde G}{\tilde G})= 4\,\sigma({\tilde G}{\tilde G}\rightarrow SS)$
and similarly for $P$. The overall factor of 4 comes from the averaging 
over initial polarisations of the goldstino in the expressions for the 
unpolarised cross section (\ref{csGGSP1}-\ref{csGGSP3}). 
Again this annihilation channel will have a negligible effect on 
determining the decoupling temperature. Other channels such as 
$S\tilde G\rightarrow P\tilde G$ will also not be important.

The form of the $S,P$ couplings to photons (\ref{SPgglag1}-\ref{SPgglag2}) 
also allows for the channel $Sf\rightarrow \gamma f$. 
A photon is exchanged in the t-channel and couples to fermion pairs 
in the standard way. In the limit $\sqrt{s} \gg m_S,m_f$ where
$m_f$ is the fermion mass one finds that
\begin{equation}
\label{csSfgf}
   \sigma(Sf\rightarrow \gamma f) = 
   {\alpha\over 2}{|{\tilde M}|^2\over F^2} 
   \left[\log\left({s^3\over m_S^4 m_f^2}\right) - {7\over 4}\right],
\end{equation} 
where $\alpha$ is the electromagnetic fine structure constant and
$f$ is a Dirac fermion (similarly for $P$ with the substitution 
$m_S\rightarrow m_P$). Notice that the dependence on the center of mass
energy $\sqrt{s}$ is now logarithmic and no longer a power law like previous 
annihilation channels. This channel (known as the Primakoff reaction)
was also considered in \cite{gmr} but in the limit $m_f\gg E_\gamma \gg 
E_S$ and is also important for establishing bounds on the gravitino mass from 
supernova cooling \cite{nd,gmr}. 
When $m_S\rightarrow 0$ the cross section becomes singular due to collinear 
singularities. In order to regulate this infrared divergence we can 
introduce a thermal mass $M_\gamma$ for the photon in the background plasma. 
Without going into a full finite temperature calculation this will be a good 
approximation to leading log order \cite{jpt}. Thus, in the massless $S$ 
limit the cross section (\ref{csSfgf}) becomes
\begin{equation}
\label{csSfgfth}
   \sigma(Sf\rightarrow \gamma f) = 
   {\alpha\over 2}{|{\tilde M}|^2\over F^2} 
   \left[\log\left({s\over M_\gamma^2}\right) - {7\over 4}\right],
\end{equation}  
and similarly for $P$. This cross section will become important in 
obtaining the decoupling temperature of the $S,P$ superpartners.

Similarly $S$ and $P$ can both scatter off the background photons via
a t and s-channel photon exchange with various particles in the final
state. The scattering cross section 
for a photon in the final state in the massless $S,P$ limit is
\begin{equation}
\label{csSggP}
    \sigma(S\gamma\rightarrow \gamma P) = \sigma(P\gamma\rightarrow \gamma S)
    = {11\over 192\pi} {|{\tilde M}|^4\over F^4} s.
\end{equation}
Similarly the cross section for pair production is
\begin{equation}
\label{csSgff}
    \sigma(S\gamma\rightarrow f\bar{f}) = \sigma(P\gamma\rightarrow f\bar{f})
    = {\alpha\over 6} {|{\tilde M}|^2\over F^2} \left(1-{4 m_f^2\over s}
    \right)^{3/2}.
\end{equation}
The process (\ref{csSggP}) is comparable to (\ref{csSSgg}), while 
(\ref{csSgff}) is subdominant compared to (\ref{csSfgfth}) and neither will 
have a large effect in the determination of the decoupling temperature.

Up to now we have implicitly neglected the possibility of cubic scalar 
couplings in the previous goldstino and scalar channels.
In general these cubic scalar interactions arise from nonminimal terms 
in the Kahler potential and superpotential terms \cite{bfzI}.
While in all realistic models the cubic scalar couplings 
vanish it is interesting to comment on their effect if the scalar annihilation
channel were to include such a term. For example, the effect of the scalar 
cubic couplings on the cross section (\ref{csSSgg}) can be seen if we 
parametrise the interaction as ${\cal L}= -A_S/3!\,S^3$.
Including this term allows for an s-channel $S$ exchange 
diagram contributing to the annihilation into photons. 
In the massless $S$ limit we obtain
\begin{equation}
\label{csSSggnew}
    \sigma(SS\rightarrow \gamma\gamma) = {1\over 64\pi} {|{\tilde M}|^2
      \over F^2} \left( 5 {|{\tilde M}|^2\over F^2} s + 4\sqrt{2} {
        {\rm Re}{\tilde M}\over F} A_S + 2 {|A_S|^2\over s} \right).
\end{equation}
The interesting feature of the above cross section is that while at
high energies we reproduce the result (\ref{csSSgg}), at low
energies the cross section may grow due to the $1/s$ behaviour. 
Cosmologically this would mean that the scalar $S$ would actually come 
back into thermal equilibrium at late times if the coupling $A_S$ were 
large enough. Similarly a cubic scalar coupling of the form $\sim S^2 P$ 
would modify (\ref{csSggP}) to a form like (\ref{csSSggnew}) via a 
u-channel scalar exchange diagram.

There are also dimension-6 terms $\sim SS F_{\mu\nu}F^{\mu\nu}$ 
(similarly $SP$, $PP$) which in a more general framework must also 
be included\footnote{We thank F. Feruglio on this point.}. The
coupling is controlled by the second derivative of the gauge kinetic 
function. Again for simplicity we will assume this term to be absent.

\section{The decoupling temperature}

The decoupling temperature of a particle species in the early universe 
is obtained in general by solving a set of coupled Boltzmann equations. 
However, in practice a reasonable estimate of the decoupling temperature 
can be obtained by checking when the interaction rate of a particular
particle species begins to fall behind the Hubble expansion rate $H$. 
The interaction rate for the scattering process $1+2\rightarrow 
{\cal F}$ into a set of final states $\cal F$ is related to the 
thermally averaged cross section times velocity, which
can be calculated using the definition \cite{gg}
\begin{equation}
        \label{tarate}
        \langle\sigma v_{M\o l}\rangle = {\int dn_1^{eq} dn_2^{eq} \sigma
        v_{M\o l}\over \int dn_1^{eq} dn_2^{eq}},
\end{equation}
where 
\begin{equation}
        \label{dn}
        dn_i^{eq} = g_i {d^3 p_i\over (2\pi)^3} f(E_i,T),
\end{equation}
$g_i$ is the number of internal degrees of freedom for the particle 
species $i$ and $f(E_i,T)$ is the statistical distribution function for
a particle with energy $E_i$ in a thermal bath at temperature $T$.
The factor $v_{M\o l}=\sqrt{(p_1\cdot p_2)^2 -m_1^2 m_2^2}/(E_1 E_2)$ is 
known as the M\o ller velocity (see Ref.~\cite{gg}) and $\sigma$ is the 
sum over all possible scattering channels of particles 1 and 2. 

In general equation (\ref{tarate}) can only be solved numerically, but
if the particle species $i$ is massless then it is possible
to obtain analytic formulae for (\ref{tarate}). Since the decoupling
temperature of a particle species $i$ will be calculated when 
$T \gg m_i$ it will be a good approximation to use 
the analytic formulae for massless particles. For a cross section
with a power law like behaviour and parametrised as $\sigma=\sum_i 
{\hat\sigma_i} s^{n_i}$, where s is the
Mandelstam variable one finds that for Re$(n_i)>-3$
\begin{equation}
\label{sivexact}
  \langle\sigma v_{M\o l}\rangle =
  {1\over\zeta(3)^2}\sum_i {C_i\over (n_i+2)} \left[\Gamma(n_i+3)
      \zeta(n_i+3)\right]^2 {\hat\sigma}_i\,T^{2n_i}, 
\end{equation}
where $C_i=2^{2n_i}-1, \left((2^{2+n_i}-1)^2/18\right)$ for Bose-Einstein 
(Fermi-Dirac) statistics, $\zeta$ is the Riemann zeta function and 
$\Gamma$ is the Euler gamma function. If the cross section has
a logarithmic dependence on $s$ then for $\sigma={\hat\sigma_0} \log s
+{\hat\sigma_1}$ with a boson and fermion in the initial state we obtain 
\begin{equation}
\label{mixedexact}
  \langle\sigma v_{M\o l}\rangle =
  {\hat\sigma}_0\log T^2+2.63327{\hat\sigma}_0 +{\hat\sigma}_1.
\end{equation}

Let us first calculate the annihilation rate for the goldstino. In the
limit that $\sqrt{s}\gg m_{3/2},m_S,m_P$ the dominant goldstino
annihilation cross section is that into photons (\ref{csGGgg}). Using the 
fact that the equilibrium number density of relativisitic goldstinos is
$n_{\tilde G} = 3 \zeta(3)/(2\pi^2) T^3$ the goldstino annihilation 
rate is \cite{tg}
\begin{equation}
\label{Grate}
    \Gamma_{\tilde G}=n_{\tilde G}\langle\sigma v_{M\o l}\rangle 
    \simeq 1.22 {{\tilde M}^2\over F^4} T^7.
\end{equation}
Similarly, we can determine the interaction rate for the scalars $S$ and 
$P$. Comparing the cross section (\ref{csSSgg}) (or (\ref{csSggP}))
with (\ref{csSfgfth}) we find that the scattering of $S,P$ off the 
background thermal plasma (\ref{csSfgfth}) will be the dominant process 
determining the interaction rate. 
For the scattering off relativistic fermions with equilibrium number 
density $n_f=3\zeta(3)/\pi^2 T^3$ the scalar interaction rate becomes
\begin{equation}
\label{Srate}
    \Gamma_{Sf}=n_f\langle\sigma v_{M\o l}\rangle 
    \simeq 0.183 \alpha {{\tilde M}^2\over F^2} T^3\left[\log\left({T^2\over
        M_\gamma^2}\right)+0.88\right],
\end{equation}
and similarly for $P$. As discussed previously, nonzero cubic scalar 
couplings can affect the interaction rate at lower temperatures via
the s-channel annihilation of scalars into two photons (\ref{csSSggnew}).
In this scenario with an equilibrium scalar number density 
$n_S= \zeta(3)/\pi^2\,T^3$ the interaction rate is given by 
\begin{equation}
\label{S3rate}
      \Gamma_{SS}=n_S\langle\sigma v_{M\o l}\rangle 
      \simeq 0.009 A_S^2 {{\tilde M}^2\over F^2} T.
\end{equation}
We will use this rate to make sure that the scalars do not
come back into thermal equilibrium during nucleosynthesis.

Notice that for temperatures $T\sim {\cal O}(100)$ MeV the scalar $S$ and
$P$ interaction rate (\ref{Srate}) is larger than the goldstino rate 
(\ref{Grate}). Consequently $S$ and $P$ will remain in thermal contact 
with the plasma much longer than the goldstinos.
This can be clearly seen when the interaction rates are compared with the
Hubble expansion rate of the universe $H$. If we express the energy density 
of the universe, $\rho(T)$ in terms of the photon energy density 
$\rho_\gamma(T)=\pi^2/15 T^4$ then the Hubble expansion becomes
\begin{equation}
\label{Hubble}
      H=\sqrt{{8\pi G_N\over 3} \rho}={\pi\over\sqrt{90}}g_\rho^{1/2}
      {T^2\over M_P},
\end{equation}
where $g_\rho$ is the effective number of relativisitic degrees of freedom.
\begin{figure}
   \centerline{
     \epsfxsize=400pt
     \epsfysize=300pt
     \hspace*{0in}\epsffile{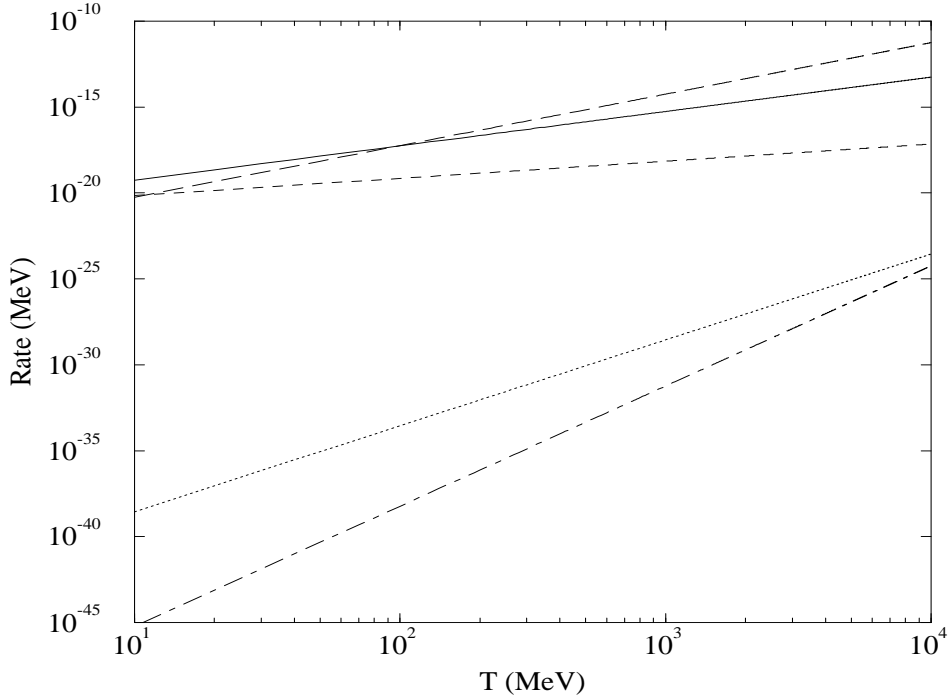}}
   \caption{\it The Hubble expansion rate (solid line), $\Gamma_{Sf}$ 
     (long-dashed line), $\Gamma_{SS}$ s-channel ($S$ exchange) 
     (dashed line), $\Gamma_{SS}$ t-channel ($\gamma$ exchange) 
     (dotted line) and $\Gamma_{\tilde G\tilde G}$ (dot-dashed line) as 
     a function of temperature in the early universe. 
     Notice that the goldstino decouples before the spin-0 superpartners.}
\label{ratesfig}
\end{figure}
In Figure~\ref{ratesfig} we have shown how the various interaction rates 
compare with the Hubble expansion rate for the representative values of 
$g_\rho=65/4, {\tilde M}=100$ GeV, $F=3.8\times10^{9}$ GeV$^2$ and we
have used $M_\gamma \sim \sqrt\alpha T$. It is clear from the figure that 
the goldstino will decouple at times
much earlier than the spin-0 scalar superpartners and that the dominant
scattering processes for $S$ and $P$ come from the scattering 
off background fermions (\ref{Srate}). Also in the figure we can see 
the linear temperature dependence of (\ref{S3rate}) (with $A_S=0.01$ GeV) 
and the possibility of $\Gamma \gtrsim H$ at lower temperatures.

Let us now obtain the expressions for the decoupling temperature.
In principle to determine the decoupling temperatures of $S,P$ and
$\tilde G$ we would need to solve a set of coupled Boltzmann equations.
However, in practice since the interaction rates are significantly different
between the scalar superpartners and the goldstino one finds that
the coupled differential equations become separate equations controlled
by ratios of $\Gamma/H$. In particular the Boltzmann equations for $S(P)$ 
separate when the dominant interaction rate is $\Gamma_{Sf} 
(\Gamma_{Pf})$ since the equations do not involve either of the other 
species. Thus it will be sufficient for our purposes to 
consider the equation $\Gamma\sim H$ for the dominant interaction rate 
$\Gamma$. When the cubic scalar couplings are included we will also
compare the interaction rate with $H$ even though the equations
may become coupled. Again this will be sufficient in order to obtain 
an order of magnitude estimate.

When the cross section has power law behaviour the decoupling 
temperature may be easily obtained analytically. One finds for the 
goldstino decoupling temperature the result
\begin{equation}
\label{Gdt}
     T_{\tilde G}\simeq 0.77 g_\rho^{1/10} \left({F^4\over{\tilde M}^2 M_P}
     \right)^{1/5}.
\end{equation}
For the spin-0 superpartners the dominant scattering mode is $\Gamma_{Sf}$.
In this case, an analytic expression for the decoupling temperature of
the spin-0 superpartners of the goldstino can be obtained since for 
a plasmon mass $M_\gamma \sim \sqrt\alpha T$ the temperature dependence
in the logarithmn cancels in $\Gamma_{Sf}$. Thus by requiring 
$\Gamma_{Sf}\sim H$ the decoupling temperature is given by
\begin{equation}
\label{Sdt}
     T_S\simeq {1.81\over(0.88-\log\alpha)} {g_\rho^{1/2}\over \alpha} 
     \left({F^2\over{\tilde M}^2 M_P}\right).
\end{equation}
Since in limit $\sqrt{s}\gg m_{3/2},m_S,m_P$ the scalars $S,P$ decouple 
later than the goldstino (see also the figure) we will use (\ref{Sdt}) to set 
a lower bound on the scale of supersymmetry breaking $F$. Note that,
without loss of generality we are also implicitly assuming that $S$ and $P$ 
decouple at the same time.

When the cubic scalar couplings are included the cross section 
(\ref{csSSggnew}) is again a power law and the analytical expression 
for the decoupling temperature is
\begin{equation}
\label{S3dt}
     T\simeq 0.03 g_\rho^{-1/2} A_S^2 \left({{\tilde M}^2 M_P\over F^2}
       \right).
\end{equation}
We will need to make this temperature as low as possible in order that the
fields $S$ (and $P$) do not come back into thermal equilibrium during
nucleosynthesis.

It is also instructive to consider the limit $m_S,m_P \gg\sqrt{s}
\gg m_{3/2}$ in which only the goldstino is assumed to be in thermal contact
with the heat bath at temperatures $T\sim {\cal O}$(MeV). As noted earlier,
in this limit the goldstino annihilation cross section into photons 
changes to the form (\ref{csGGggC}). Now the annihilation cross section
into photons and fermions are comparable and qualitatively similar
$\sigma \propto s^3$. Including the electron, muon and 3 families of 
massless neutrinos for the fermions gives for the decoupling temperature
\begin{equation}
\label{GdtC}
     T_{\tilde G}\simeq 0.53 g_\rho^{1/14} \left({F^4\over M_P}
       \right)^{1/7}.
\end{equation}
This equation will be used to show how much weaker the bound on the
supersymmetry breaking scale $F$, becomes in the massive $m_S,m_P$ case.

In order to bound the decoupling temperature of a particle species $i$, 
we need to consider its effect on nucleosynthesis. This can be done
by noting that the energy density of new massless particles is equivalent 
to an effective number $\Delta N_\nu$ of additional doublet neutrinos
\begin{equation}
\label{DN}
     \Delta N_\nu = f_{B,F} \sum_i {g_i\over 2} \left[{g_\rho(T_\nu)
     \over g_\rho(T_{D_i})}\right]^{4/3},
\end{equation}
where $f_B=8/7$ for bosons, $f_F=1$ for fermions and $g_i$ is the
number of internal degrees of freedom of the particle species $i$
\cite{ss}.
If the scalars decouple at $T_{S,P} = T_\nu \simeq {\cal O}$(1 MeV) then 
according to (\ref{DN}) $S$ and $P$ will effectively behave
as 1.14 extra neutrino doublets. If, however the scalars were
to decouple during the epoch $T_\nu< T_{S,P} < T_{\mu}$, where
$T_\mu$ is the muon decoupling temperature 
then the effective number of additional doublet neutrinos during 
nucleosynthesis due to $S$ and $P$ is 0.91. While neither of these bounds
can be strictly ruled out \cite{ssII} we will nevertheless
suppose that $S$ and $P$ decouple before the muon decouples\footnote{
This is somewhat plausible given that there may be other contributions to 
$\Delta N_\nu$ arising from massive neutrinos.} to quote a
bound on the supersymmetry breaking scale $F$.

As alluded to earlier there is no loss of generality if we calculate the 
bounds using the expressions for the scalar field $S$. The bounds are 
identical if we were to use the expressions for $P$, since to leading
order the dominant cross sections have no dependence on $m_S$ or $m_P$. 
Thus using the expression (\ref{Sdt}) and requiring that $T_S > T_\mu
\simeq {\cal O}$(100 MeV) leads to the bound
\begin{equation}
\label{Sbound}
     F\gtrsim 4\times10^9 \,{\rm GeV}^2 \left({{\tilde M}\over 100\,{\rm GeV}}
     \right).
\end{equation}
This strengthens the bound considered in \cite{tg}, where the effects of 
massless $S$ and $P$ during nucleosynthesis were not considered. 
For ${\tilde M}\simeq 100$ GeV the bound on the supersymmetry
breaking scale becomes $F\gtrsim 4\times 10^{9}$ GeV$^2$. 
Using the relation $F=\sqrt{3} m_{3/2} M_p$ this leads to a bound on the 
gravitino mass $m_{3/2} \gtrsim 1$ eV. This is similar to the
bound obtained by \cite{gmr} where the logarithmic term was neglected.
One may have expected a large enhancement from the logarithmn 
factor in the massless limit due to the singularity, but since the 
plasmon mass is large, the enhancement in the cross section is only 
$\log\alpha\sim -5$ and thus impotent. Also if the mass of the scalars 
$m_{S,P}\gtrsim 1$ MeV then the bounds calculated here would be void 
because the cross section limits are no longer valid and a more detailed 
analysis in this intermediate mass range would need to be done.

When the cubic scalar couplings, represented by $A_S$ are included we need
to make sure that they do not return to thermal equilibrium during 
nucleosynthesis. Requiring $T\lesssim \cal O$(eV) leads to a bound
of $m_{3/2}\gtrsim 10^{-4} A_S$, which vanishes in the limit $A_S=0$.
We emphasise that while this bound is approximate, a more detailed 
analysis than that given here could provide interesting constraints
on theories with nonzero cubic scalar couplings.

Finally, the gravitino mass bound coming from (\ref{GdtC}) in the
limit $m_S,m_P\gg \sqrt{s} \gg m_{3/2}$ is $m_{3/2}\gtrsim 4 \times 
10^{-7}$ eV and is independent of the gaugino mass $\tilde M$. 
As expected the bound becomes much weaker in this case and there are 
already stronger bounds from collider phenomenology in this limit 
\cite{bfzIII}.

\section{Conclusion}

We have shown that in the massless limit the scalar superpartners $S,P$
of the goldstino couple much more strongly to the background thermal 
plasma than the goldstino. This means that during nucleosynthesis the 
scalars $S$ and $P$ may contribute as extra effective neutrino doublets,
while the goldstinos have long since decoupled. Requiring that $S$ and $P$ 
decouple sufficiently early, so as not to upset current nucleosynthesis
bounds, sets a lower bound on the scale of supersymmetry breaking which
depends on the gaugino mass $\tilde M$. The bound on the 
supersymmetry breaking scale assuming $\tilde M=100$ GeV is
$\sqrt{F}\gtrsim 60$ TeV which leads to a lower bound on the gravitino
mass of $m_{3/2}\gtrsim 1$ eV. These bounds are much stronger than those 
considered in \cite{tg}, but complement the cosmological bounds in
\cite{gmr}. Finally in the massive $S,P$ case the nucleosynthesis
bound is considerably weaker and similar to other recent cosmological
bounds \cite{gmt,dmt}.

\section*{Acknowledgements}
\noindent
We would like to thank F. Feruglio, G. Jungman and A. Riotto
for useful discussions.

\vfil\eject

\end{document}